\documentclass[journal]{IEEEtran}
\usepackage{amsfonts,amssymb,amsmath,bm}
\usepackage{float}
\usepackage{graphicx}
\usepackage{subfigure}
\usepackage{cite,color}
\usepackage{cases}
\usepackage{epsfig}
\usepackage{verbatim}
\usepackage{algorithm,algorithmic}
\usepackage{dblfloatfix}
\usepackage{setspace}

\IEEEoverridecommandlockouts

\newtheorem{theorem}{\bf{Theorem}}[section]
\newtheorem{lemma}{\textbf{Lemma}}

\newtheorem{remark}{\textbf{Remark}}[section]

\begin{document}
\title{\Large{Stochastic Hybrid Combining Design for Quantized Massive MIMO Systems}}
\author{Yalin Wang, Xihan Chen, Yunlong Cai, and Lajos Hanzo
  \thanks{
  Copyright (c) 2015 IEEE. Personal use of this material is permitted. However, permission to use this material for any other purposes must be obtained from the IEEE by sending a request to pubs-permissions@ieee.org.

  The work of Y. Cai was supported in part by
the National Natural Science Foundation of China under Grants 61831004 and
61971376,  the Zhejiang Provincial Natural Science Foundation
for Distinguished Young Scholars under Grant LR19F010002,    the National Key R\&D Program of China (No.2020YFB1805005), and
the State Key Laboratory of Rail Traffic Control and Safety (Contract No. RCS2020K010), Beijing Jiaotong University.
(Correspondence authors: Xihan Chen; Yunlong Cai.)

  Y. Wang and X. Chen  are with the College of ISEE, Zhejiang University, Hangzhou 310027, China (e-mail: wang\_yalin@zju.edu.cn; chenxihan@zju.edu.cn).
   Y. Cai is with the College of ISEE, Zhejiang University, Hangzhou 310027, China, and also with the State Key Laboratory of Rail Traffic Control and Safety, Beijing Jiaotong University, Beijing 100044, China (e-mail: ylcai@zju.edu.cn).
  L. Hanzo is with the Department of ECS, University of Southampton, Southampton SO17 1BJ,  U.K. (e-mail: lh@ecs.soton.ac.uk).}
  }

	\linespread{0.85}
	\maketitle
	\pagestyle{empty}
	\thispagestyle{empty}
\begin{abstract}
Both the power-dissipation and cost of massive multiple-input multiple-output (mMIMO) systems may be substantially reduced by using low-resolution analog-to-digital converters (LADCs) at the receivers. However, both the coarse quantization of LADCs           and the inaccurate instantaneous channel state information (ICSI) degrade the performance of quantized mMIMO systems. To overcome these challenges, we propose a novel stochastic hybrid analog-digital combiner (SHC) scheme for adapting the hybrid combiner to the long-term statistics of the channel state information (SCSI). We seek to minimize the transmit power by jointly optimizing the SHC subject to average rate constraints. For the sake of solving the resultant nonconvex stochastic optimization problem, we           develop a relaxed stochastic successive convex approximation (RSSCA) algorithm. Simulations are carried out to confirm the benefits of our proposed scheme over the benchmarkers.
\end{abstract}
\maketitle
\IEEEpeerreviewmaketitle
\vspace{-3.5mm}
\section{Introduction}
\vspace{-1.5mm}
Massive multiple-input multiple-output (mMIMO) systems constitute
promising techniques for next generation wireless
communication~\cite{b1}. The number of antennas is scaled up compared
to traditional MIMO systems with the objective of improving both the
energy efficiency (EE) and spectral efficiency (SE) of communications,
albeit at the cost of increasing both the cost power-dissipation of
the signal-processing and RF hardware. As a remedy, the hybrid
analog-digital combining aims for mitigating these issues by employing
fewer radio frequency (RF) chains linked to a large number of antennas
by an analog combiner~\cite{hybrid_combiner}. Since the
power-dissipation of analog-to-digital converters (ADCs) is dominated
by the RF circuit power, which scales exponentially with the number of
quantization bits, employing low-resolution ADCs (LADCs) has emerged
as a natural solution~\cite{survey1}. Therefore, the seamless
integration of hybrid combining and LADCs is of paramount importance.

Significant efforts have been made to study the feasibility of hybrid
analog-digital combining aided quantized mMIMO systems relying on
LADCs. Specifically, the authors of~\cite{survey1} developed hybrid
combiners by minimizing the mean-squared error of received signals in
the presence of multiuser interference. However, their analog combiner
design does not satisfy the constant modulus constraint, hence its
implementation is impractical. To circumvent this difficulty, the
authors of~\cite{b2} conceived a more practical hybrid combiner
relying on a partially-connected structure and rigorously derived the
achievable rate considering the correlation of quantization errors.
To further mitigate the effect of quantization errors, a two-stage
analog combiner was developed in~\cite{two_stage} for maximizing the
mutual information between the quantized and the transmitted
signals. Nevertheless, the aforementioned studies merely optimize the
hybrid combiner in a separate manner, which inevitably leads to
performance degradation. To mitigate this impediment, the authors
of~\cite{shl} harnessed fractional programming techniques for EE
maximization by opting for the joint hybrid combiner design principle.

\vspace{-1mm}
However, to the best of our knowledge, the studies in the literature
mainly depend on the knowledge of the instantaneous channel state
information (ICSI). In practice, acquiring ICSI is quite challenging
in mMIMO systems. Owing to the limited coherence time associated with
a high number of antennas, an excessive number of pilot symbols is
necessitated for accurate channel estimation in the mMIMO regime. A
further problem is the severe nonlinear distortion due to the coarse
quantization by LADCs. Fortunately, the base station (BS) is capable
of acquiring the slowly-varying statistical channel state information
(SCSI) with the aid of its long-term feedback~\cite{scsi}. Therefore,
it is more reasonable to consider a hybrid combining scheme purely
relying on the knowledge of SCSI. Moreover, the pioneering
contribution of \cite{EESE1} intimates the underlying EE vs SE
trade-off. On the other hand, most of the existing studies such
as~\cite{shl} and \cite{ee1} only place particular emphasis on
improving the EE of quantized mMIMO systems. As such, how to design
hybrid combining schemes based on the knowledge of SCSI for quantized
mMIMO systems when the trade-off between the EE and SE is explicitly
considered deserves further study.

\setcounter{equation}{4}
\begin{figure*}[b]
\hrulefill
\begin{equation}\label{sinr}
\mathrm{SINR}_{k} = \frac{p_{k}\left|\mathbf{w}_{k}^{H}\mathbf{V}^{H}\mathbf{Q}_{\gamma}\mathbf{U}^{H}\mathbf{h}_{k}\right|^{2}}{\sum_{i\neq k}p_{i}\left|\mathbf{w}_{k}^{H}\mathbf{V}^{H}\mathbf{Q}_{\gamma}\mathbf{U}^{H}\mathbf{h}_{i}\right|^{2} + \sigma^{2}\left\|\mathbf{w}_{k}^{H}\mathbf{V}^{H}\mathbf{Q}_{\gamma}\mathbf{U}^{H}\right\|^{2} + \mathbf{w}_{k}^{H}\mathbf{V}^{H}\mathbf{R}_{q}\mathbf{V}\mathbf{w}_{k}}
\end{equation}
\end{figure*}
\setcounter{equation}{0}
\vspace{-1mm}
To shed more light on these critical issues, we devise a
novel stochastic hybrid combiner (SHC) scheme for quantized mMIMO
systems and assume that only the SCSI is available at the BS. By
invoking beamspace mMIMO techniques which steer the arriving signals
having various angles of arrival to distinct array elements, we can
significantly reduce the number of RF chains and conceive
cost-efficient implementations. Our interest in this compact paper lies in striking a compelling throughput vs. power consumption trade-off, namely a SE vs. EE trade-off. To this end,
we seek to minimize the transmit power by jointly optimizing the SHC
scheme subject to average rate constraints. For efficiently solving
this nonconvex stochastic constrained optimization problem, we propose
a \emph{relaxed stochastic successive convex approximation} (RSSCA)
algorithm, which intrinsically amalgamates a binary relaxation
technique with a stochastic successive convex approximation (SSCA)
solution. The proposed SHC scheme only adapts the hybrid combiner to
the long-term SCSI and offers compelling advantages over the existing
schemes.  Firstly, as a benefit of the \emph{channel hardening}
phenomenon of mMIMO systems~\cite{hardening}, a fading channel behaves
almost deterministically. Hence, considering the hybrid combiner
relying on SCSI is more practical and efficient, yet without
substantial performance degradation. Secondly, the hybrid combiner is
only updated with the aid of the outdated CSI samples, hence
mitigating the CSI signaling latency. Our simulation results validate
that the proposed scheme outperforms the benchmarkers.
	

\vspace{-2.5mm}
\section{System Model and Problem Formulation}
\vspace{-1mm}
\setcounter{equation}{7}
\begin{figure*}[b]\small
\hrulefill
\begin{equation}\label{chi}
\chi= \left\{\bm{x}: \bm{p}\in[0,P_{k}^{\mathrm{max}}]^{K}; \sum_{i=1}^{N}c_{ij} = 1; \sum_{j=1}^{S}c_{ij}\leq 1; c_{ij} \in \{0,1\},\forall{i,j} \right\}
\end{equation}
\end{figure*}
\setcounter{equation}{0}
\begin{figure}[ht]
    \centering
    \includegraphics[width=8cm]{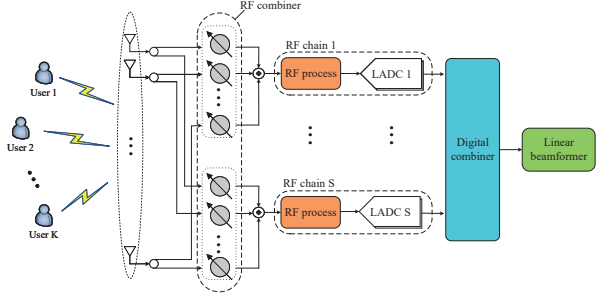}\\
    \caption{An illustration of the SHC scheme in a multi-user mMIMO uplink system with LADCs.}\label{fig:ladc_system}
    \end{figure}
\subsection{Network Architecture and Frame Structure}
\vspace{-2mm} We consider a single-cell multiuser quantized mMIMO
uplink system, where a BS supports $K$ users. Each user is equipped
with a single antenna and the BS is equipped with $M > 1$ antennas and
$S \ll M$ RF chains. We focus on a fully-connected RF combining
structure at the BS, where each RF chain is connected to all receive
antennas using phase shifters~\cite{binary} and LADCs
are employed between the RF combiner and the digital combiner for
converting the analog signals to discrete-amplitude signals. To enable the
reliable retrieval of data symbols for each user, a low-complexity linear
receive beamformer
$\mathbf{w}_{k}\in\mathbb{C}^{S\times 1}$ is also employed at the
BS. Specifically, the received signal vector is combined by a
hybrid combiner using LADCs, as shown in Fig. \ref{fig:ladc_system}.

Let $p_{k}$ be the transmit power of
user $k$, $\mathbf{h}_{k}\in\mathbb{C}^{M\times 1}$ be the uplink
channel spanning from user $k$ to the BS, $s_{k}\sim\mathcal{CN}(0,1)$ be the
transmitted data of user $k$, and $\mathbf{n}\in\mathbb{C}^{M\times 1}$
be the additive white Gaussian noise (AWGN) at the BS with
distribution $\mathcal {C}\mathcal
{N}(0,{\sigma^2}\mathbf{I}_M)$. Assuming a narrowband channel, the
signal received at the BS can be expressed as
\begin{equation}
\mathbf{y} =\sum_{k=1}^{K}\sqrt{p_{k}}\mathbf{h}_{k}s_{k} + \mathbf{n}= \mathbf{H}\mathbf{P}^{\frac12}\mathbf{s}+\mathbf{n},
\end{equation}
where $\mathbf{P} = \mathrm{diag}(p_{1},\cdots,p_{k})$, $\mathbf{H} = [\mathbf{h}_{1},\cdots,\mathbf{h}_{K}]\in\mathbb{C}^{M\times K}$, and $\mathbf{s} = [s_{1},\cdots,s_{k}]^{T}$.

The received signal $\mathbf{y}$ is first combined by the RF combiner
$\mathbf{U}\in\mathbb{C}^{M\times S}$, which is implemented based on a
discrete Fourier transform (DFT) codebook $\mathbf{D}$. Consequently,
such a DFT codebook-based RF combiner can be formulated as
$\mathbf{U} = \mathbf{D}\mathbf{C}$, where $\mathbf{D} =
[\mathbf{d}_{1},\cdots,\mathbf{d}_{N}]\in\mathbb{C}^{M\times N}$
denotes the codebook of size $N$ and $\mathbf{C}\in\mathbb{C}^{N\times
  S}$ denotes the selection matrix with element $c_{ij}\in\{0,1\}$ for
selecting codewords. It can be readily implemented utilizing a static
phase shifting network in the RF domain along with an RF
switch. Therefore, the output of the RF combiner can be expressed as
$
\tilde{\mathbf{y}} = \mathbf{U}^{H}(\mathbf{H}\mathbf{P}^{\frac{1}{2}}\mathbf{s}+\mathbf{n}).
$
Each of the RF combining output $\tilde{\mathbf{y}}$ is connected to a
LADC pair, which separately quantizes the imaginary and real part of the
signal $\tilde{\mathbf{y}}$. Considering that each of the LADC has $q$
quantization bits, the quantized output under the additive
quantization noise model (AQNM)~\cite{shl} is expressed as
\begin{equation}
\mathbf{y}_{q} \!=\! f(\tilde{\mathbf{y}}) = \mathbf{Q}_{\gamma}\tilde{\mathbf{y}} + \mathbf{n}_{q} = \mathbf{Q}_{\gamma}\mathbf{U}^{H}\mathbf{H}\mathbf{P}^{\frac{1}{2}}\mathbf{s} + \mathbf{Q}_{\gamma}\mathbf{U}^{H}\mathbf{n} + \mathbf{n}_{q},
\vspace{-1mm}
\end{equation}
where $f(\cdot)$ is the element-wise quantization function,
$\mathbf{Q}_{\gamma} =
\mathrm{diag}(\gamma,\cdots,\gamma)\in\mathbb{C}^{S\times S}$ with
quantization gain $\gamma=1 - \rho$, and $\rho$ is a normalized
quantization error. For $\rho\leq5$, the typical
  values of $\rho$ are listed in \cite{bit}, while for $\rho\geq5$,
  they can be approximated by $\rho = \frac{\pi\sqrt3}{2}2^{-2q}$.
The additive quantization noise $\mathbf{n}_{q}$ obeys the complex-valued
Gaussian distribution with zero mean so that $\mathbf{n}_{q}$ and
$\tilde{\mathbf{y}}$ are uncorrelated with each other.
For a fixed channel realization $\mathbf{H}$, the covariance matrix of $\mathbf{n}_{q}$ is written as
$\mathbf{R}_{q} = \mathbb{E}[\mathbf{n}_{q}\mathbf{n}_{q}^{H}] = \mathbf{Q}_{\gamma}(\mathbf{I}-\mathbf{Q}_{\gamma})\mathrm{Diag}(\mathbf{U}^{H}\mathbf{H}\mathbf{P}\mathbf{H}^{H}\mathbf{U} + \sigma^2\mathbf{U}^{H}\mathbf{U})$.
Once the received signals are quantized, the baseband combiner
$\mathbf{V}\in\mathbb{C}^{S\times S}$ at the BS is applied to handle
the quantization noise introduced by the LADCs, additionally mitigating
the multiuser interference. The corresponding output is given by
\begin{equation}
\bar{\mathbf{y}} \!=\! \mathbf{V}^{H}{\mathbf{y}_{q}} = \mathbf{V}^{H}\mathbf{Q}_{\gamma}\mathbf{U}^{H}\mathbf{H}\mathbf{P}^{\frac{1}{2}}\mathbf{s} + \mathbf{V}^{H}\mathbf{Q}_{\gamma}\mathbf{U}^{H}\mathbf{n} + \mathbf{V}^{H}\mathbf{n}_{q}.
\vspace{-1mm}
\end{equation}
Finally, the retrieved data symbol of user $k$ after
the low-complexity linear receive beamformer can be
expressed as
\begin{equation}
\hat{s_{k}}\!=\!\mathbf{w}_{k}^{H}\mathbf{V}^{H}\mathbf{Q}_{\gamma}\mathbf{U}^{H}\mathbf{H}\mathbf{P}^{\frac{1}{2}}\mathbf{s} \!+\! \mathbf{w}_{k}^{H}\mathbf{V}^{H}\mathbf{Q}_{\gamma}\mathbf{U}^{H}\mathbf{n} \!+\! \mathbf{w}_{k}^{H}\mathbf{V}^{H}\mathbf{n}_{q}.
\vspace{-1mm}
\end{equation}

By referring to the system model, the instantaneous rate of
user $k$ can be explicitly expressed as
$r_{k}(\mathbf{P},\mathbf{C},\mathbf{V},\mathbf{W}) = \log(1+\mathrm{SINR}_{k})$,
where $\mathbf{W} = [\mathbf{w}_{1},\cdots,\mathbf{w}_{K}]\in\mathbb{C}^{S\times K}$ is the composite receive beamforming vector and $\mathrm{SINR}_{k}$ denotes the received signal-to-interference-plus-noise ratio (SINR) of user $k$ defined in (\ref{sinr}), shown at the bottom of this page.
Hence, the average rate of user $k$ is given by
$\bar{r}_{k}(\mathbf{P},\mathbf{C},\mathbf{V},\mathbf{W}) = \mathbb{E}[r_{k}(\mathbf{P},\mathbf{C},\mathbf{V},\mathbf{W})]$.



\begin{remark}
In contrast to the conventional hybrid combiner structure based on the
knowledge of ICSI, both the RF combiner $\mathbf{U}$ and the baseband
combiner $\mathbf{V}$ are only adapted to the SCSI in our proposed SHC
scheme. By exploiting the \emph{channel hardening} property of
quantized mMIMO systems, the gain of adapting the power allocation
based on the ICSI remains modest~\cite{power}. Hence, we adapt the
power allocation strategy $\mathbf{P}$ to the long-term SCSI in
consideration of the signaling overhead.
\end{remark}

\vspace{-3.5mm}
\subsection{Frame Structure}
\vspace{-2mm}
In the proposed SHC scheme, we divide the time domain into several
super-frames and each super-frame consists of $L_{f}$ frames. We
further divide each frame into $L_{s}$ time slots, where the channel
remains constant within each time slot. Thanks to the advanced
compressive sensing based channel estimation methods, it is possible
to achieve efficient uplink training with the aid of a limited number
of RF chains~\cite{binary}. Under this setup, we can obtain one
(possibly outdated) channel sample $\mathbf{H}$ at each frame. To be
more specific, each user sends dedicated uplink pilots to the BS, and
the effective channel gains are subsequently estimated based on the
received pilot signals. It is noteworthy that the variables
$\mathbf{P},\mathbf{C},\mathbf{V},\mathbf{W}$ are only updated once
based on a single channel sample in each frame to achieve a mMIMO array gain
at a reduced implementation cost.
\vspace{-4mm}
\subsection{Problem Formulation}
\vspace{-2mm}
We are interested in designing a hybrid combiner for the uplink of
quantized mMIMO systems for minimizing the system's power-dissipation
subject to the average rate requirement of each user. In practical
implementations power control is of pivotal importance, especially
in the uplink of quantized mMIMO system. Upon reducing the transmit
power, the battery life of power constrained devices will be
commensurately prolonged. In particular, the DFT-based RF combiner is
adopted in the SHC for striking a performance vs. hardware cost
trade-off, where the choice of codewords must strictly meet the
following two criteria: 1) each RF chain is associated with a single
codeword, 2) each codeword is assigned to no more than one RF
chain. We consider the average rate as our QoS metric and denote the
target rate requirement of user $k$ by $\gamma_{k} > 0$. More formally, the
problem can be formulated as
\setcounter{equation}{5}
\begin{subequations} \label{P}
\begin{align}
\underset{\mathbf{P},\mathbf{C},\mathbf{V},\mathbf{W}}{\min}\quad&\sum\nolimits_{k=1}^{K}p_{k}\label{P1}\\
\mbox{s.t.}\quad
&\bar{r}_{k}(\mathbf{P},\mathbf{C},\mathbf{V},\mathbf{W}) \geq \gamma_{k},\quad\forall{k}, \label{P1A}\\
&p_{k} \leq P_{k}^{\mathrm{max}},\quad\forall{k}, \label{P1B}\\
&\sum_{i=1}^{N}c_{ij} = 1,\quad\forall{j}, \label{P1C}\\
&\sum_{j=1}^{S}c_{ij}\leq 1,\quad\forall{i},  \label{P1D}\\
&c_{ij} \in \{0,1\},\quad\forall{i,j}, \label{P1E}
\end{align}
\end{subequations}
where (\ref{P1A}) is the average rate requirement and (\ref{P1B}) is
the transmit power constraint of each user. The constraints (\ref{P1C})
and (\ref{P1D}) guarantee the realization of codewords selection
criteria.

\vspace{-5mm}
\section{Stochastic Hybrid Combining Scheme}
\vspace{-1.5mm}
There are three major challenges in solving problem (\ref{P}): i) the
nonconvexity of the constraint functions, mainly due to the coupling
variables in constraint (\ref{P1A}) and the discrete binary variable
$c_{ij}$ in (\ref{P1E}); ii) the NP-hard property of the mixed integer
nonlinear programming caused by (\ref{P1E}); and iii) the stochastic
nature of constraint (\ref{P1A}). In the sequel, we develop an
efficient algorithm to address this problem iteratively.
\vspace{-3.5mm}
\subsection{Problem Transformation and Surrogate Function }
\vspace{-1.5mm}
To tackle the difficulty arising from the discrete feasible region, we
replace constraint (\ref{P1E}) that $c_{ij}$ be 0 or 1 with the
relaxed constraint that it be in the interval [0,1]. To obtain an
integer solution for the codeword selection indicator, we rely on
the method of~\cite{binary} to round each $\hat{c}_{ij}$ generated
by the proposed algorithm to the nearest integer as follows.
\begin{equation}
c_{ij} = \left\{
\begin{aligned}
&\lfloor{\hat{c}_{ij}}\rfloor=0,\quad \text{if}\quad \hat{c}_{ij}- \lfloor{\hat{c}_{ij}}\rfloor \leq \varepsilon_{j},\\
&\lceil \hat{c}_{ij}\rceil=1,\quad \text{otherwise}, \\
\end{aligned}
\right.\forall {i,j},
\vspace{-1mm}
\end{equation}
where $0\leq\varepsilon_{j} \leq1$ is chosen using a simple bisection method, so that both the constraints~\eqref{P1C} and \eqref{P1D} are met.

For convenience, we let  $\mathbf{p} = \mathrm{diag}(\mathbf{P})$, $\mathbf{c} = \mathrm{vec}(\mathbf{C})$, $\mathbf{v} = \mathrm{vec}(\mathbf{V})$, $\mathbf{w} = \mathrm{vec}(\mathbf{W})$ and define variable $\mathbf{x}\triangleq[\mathbf{p}^{T},\mathbf{c}^{T},\mathbf{v}^{T},\mathbf{w}^{T}]^{T}\in \chi$, where $\chi$ is a convex constraint set defined in (\ref{chi}) as displayed at the bottom of this page. The key observation is that $\chi$ has a decoupled form: $\chi = \{\mathbf{x}: x_{t}\in {\chi}_{t}, t = 1,\ldots,n\}$, where ${\chi}_{t}$ is a convex region in $\mathbb{C}$. Then constraint (\ref{P1A}) can be rewritten as
$f_{k}(\mathbf{x}) \triangleq \gamma_{k} - \bar{r}_{k}(\mathbf{x})\leq 0 \label{stochastic constraint}.$
It is challenging to accurately calculate the expectations in the
constraint function $f_{k}(\mathbf{x})$. To handle
such a stochastic constraint, we first construct the quadratic
surrogate function $\hat{f}_{k}^{l}(\mathbf{x})$ of the constraint
function $f_{k}(\mathbf{x})$, which efficiently handles the
expectations and facilitates rapidly-converging algorithm design at
a low complexity. $\hat{f}_{k}^{l}(\mathbf{x})$ can be recognized as
a convex approximation of $f_{k}(\mathbf{x})$ at each iteration.
Each channel sample corresponds to an iteration. Let $\mathbf{x}^{l}$
denote the variable used during the $l$-th channel
sample. Specifically, $\hat{f}_{k}^{l}(\mathbf{x})$ is formulated as
\setcounter{equation}{8}
\begin{equation} \label{surrogate}
\hat{f}_{k}^{l}(\mathbf{x}) = \gamma_{k} - \hat{r}_{k}^{l} + \Re[(\boldsymbol{\kappa}_{k}^{l})^{H}(\mathbf{x} - \mathbf{x}^{l})] + \tau_{k}\|\mathbf{x} - \mathbf{x}^{l}\|^{2},
\vspace{-1mm}
\end{equation}
where $\tau_{k}$ is a positive constant so that it ensures the strong convexity of $\hat{f}_{k}^{l}(\mathbf{x})$. For given channel samples $\mathbf{H}^{i}$,
$\hat{r}_{k}^{l} = \sum_{i}^{l}r_{k}(\mathbf{x}^{l}; \mathbf{H}^{i})/l, \forall i = 1,\ldots,l,$
represents the sample average approximations for $\bar{r}_{k}(\mathbf{x}^{l})$, and $\boldsymbol{\kappa}_{k}^{l}$ is an approximation for the gradient $\nabla f_{k}(\mathbf{x}^{l})$, which is updated recursively as
\begin{equation}
\boldsymbol{\kappa}_{k}^{l} = (1 - \beta^{l}){\boldsymbol\kappa}_{k}^{l-1} + \beta^{l}\boldsymbol{\eta}_{k}(\mathbf{x}^{l}),
\vspace{-1mm}
\end{equation}
with $\boldsymbol{\kappa}_{k}^{-1} = \boldsymbol{0}$, where
$\beta^{l}\in(0,1]$ is the step-size sequence and
    $\boldsymbol{\eta}_{k}$ is the gradient of the instantaneous rate
    $r_{k}(\mathbf{x})$ w.r.t. $\mathbf{x}$. Then
    $\boldsymbol{\eta}_{k}$ can be written as
\begin{equation}
\boldsymbol{\eta}_{k}(\mathbf{x}) = [\nabla_{\mathbf{p}}^{T} r_{k}, \nabla_{\mathbf{c}}^{T} r_{k}, \nabla_{\mathbf{v}}^{T} r_{k},\nabla_{\mathbf{w}}^{T} r_{k}]^{T},
\vspace{-1mm}
\end{equation}
where $\nabla_{\mathbf{p}} r_{k}, \nabla_{\mathbf{c}} r_{k}, \nabla_{\mathbf{v}} r_{k}, \nabla_{\mathbf{w}} r_{k}$ are the gradients of $r_{k}$ w.r.t. $\mathbf{p},\mathbf{c},\mathbf{v}, \mathbf{w}$, respectively, based on the matrix calculus and the chain rule.



\setcounter{equation}{14}
\begin{figure*}[b]
\hrulefill
\footnotesize
\begin{align}\label{lagrange}
\mathcal{L}^{l}(\mathbf{x},\Theta) &= \sum_{k=1}^{K}p_{k} + \sum_{k=1}^{K}\lambda_{k}\hat{f}_{k}^{l} + \sum_{i=1}^{N}\mu_{i}(\sum_{j=1}^{S}c_{ij} - 1) + \sum_{k=1}^{K}\varrho_{k}(p_{k} - P_{k}^{\mathrm{max}}) + \sum_{j=1}^{S}\delta_{j}(\sum_{i=1}^{N}c_{ij} - 1) + \sum_{i=1}^{N}\sum_{j=1}^{S}\phi_{ij}(c_{ij}-1)\nonumber \\
&=\sum_{t=1}^{n}a(\Theta)|x_{t}|^{2} + \Re\left[\sum_{t=1}^{n}b(\Theta)x_{t}\right]  +c(\Theta)
\end{align}
\end{figure*}
\vspace{-4mm}
\subsection{Proposed RSSCA Algorithm}
\vspace{-2mm}
The proposed RSSCA algorithm iteratively minimizes a sequence of
surrogate functions. Recalling~\eqref{surrogate} allows us to express
the problem (\ref{P}) as the following approximated
convex one
\setcounter{equation}{11}
\begin{equation}\label{P2}
\underset{\mathbf{x}\in\chi}{\min}\quad\sum\nolimits_{k=1}^{K}p_{k} 
\quad \mbox{s.t.}\quad
\hat{f}_{k}^{l}(\mathbf{x})\leq 0,\quad\forall{k}. 
\vspace{-1mm}
\end{equation}
The details of the solution of problem (\ref{P2}) will be postponed to
Section III-C. Note that problem (\ref{P2}) may not have a feasible
optimal solution. If problem (\ref{P2}) turns out to be infeasible, we
have to construct the following feasibility problem.
\begin{equation} \label{P3}
\underset{\mathbf{x}\in\chi,\xi}{\min}\quad\xi 
\quad \mbox{s.t.}\quad
\hat{f}_{k}^{l}(\mathbf{x})\leq \xi,\quad\forall{k}, 
\end{equation}
which can be efficiently solved by the convex programming toolbox
CVX. The solution of problem~(\ref{P3}) can be recognized as the
projection of problem~(\ref{P2}) onto the point, which is the closest
to the feasible region of problem~\eqref{P2}.

Given the optimal solution $\bar{\bm{x}}^l$ in problem (\ref{P2}) or (\ref{P3}), $\mathbf{x}$ is updated as
\begin{equation} \label{update_x}
\mathbf{x}^{l+1} = (1-\alpha^{l})\mathbf{x}^{l}+\alpha^{l}\bar{\mathbf{x}}^{l},
\vspace{-1mm}
\end{equation}
where $\alpha^{l}\in(0,1]$ is the step-size sequence. Then the above
iteration is carried out until convergence is reached. We summarize
the details of the proposed RSSCA in Algorithm~1,
where $\mathrm{unvec}_{n,m}(\cdot)$ represents
the operation, which turns the $nm\times 1$ column
vector into a matrix of size $n\times m$.

\begin{algorithm}[t]
\small
	\caption{Proposed RSSCA Algorithm\label{algorithm}}
	\textbf{Initialization:} $L_{f}$; $\{\alpha^{l}\},\{\beta^{l}\}$; $\mathbf{x}^0 \in \chi$; $\boldsymbol{\kappa}_{k}^{-1} = \boldsymbol{0},\forall k$; $l=0$.\\
    \textbf{Step 1:} {Obtain a channel sample $\mathbf{H}^{l}$ within frame $l$}.

    \qquad\quad\;\,{Update the surrogate function $\hat{f}_{k}^{l}(\mathbf{x}),\forall k$ using (\ref{surrogate})}.\\
	\textbf{Step 2:} {Solve (\ref{P3}) to obtain the optimal solution $\hat{\xi},\hat{\mathbf{x}}.$}\\
    \textbf{If} $\hat{\xi}\leq 0 \textbf{:}$ {Solve ($\ref{P2}$) to obtain $\bar{\mathbf{x}}^{l}$}. (Problem ($\ref{P2}$) is feasible)\\
    \textbf{Else:} {Let $\bar{\mathbf{x}}^{l} = \hat{\mathbf{x}}.$} \textbf{End if}\\
    \textbf{Step 3:} {Update {$\mathbf{x}^{l+1}$ according to (\ref{update_x})}}.\\
    \textbf{Step 4:} {Let $l = l+1$. If $l \neq L_{f}$, return to Step 1. Otherwise, terminate the algorithm.}\\
    \textbf{Output:} {$\mathbf{P}=\mathrm{diag}(\mathbf{x}^{l}[1:K])$; $\mathbf{C}=\mathrm{unvec}_{N,S}(\mathbf{x}^{l}[K+1:K+NS])$; $\mathbf{V}=\mathrm{unvec}_{S,S}(\mathbf{x}^{l}[K+NS+1:K+NS+S^{2}])$; $\mathbf{W}=\mathrm{unvec}_{S,K}(\mathbf{x}^{l}[K+NS+S^{2}+1:K+NS+S^{2}+SK])$}.

\end{algorithm}

\vspace{-5mm}
\subsection{Solutions for Problem (\ref{P2})}
\vspace{-2.5mm}
We provide solutions to the quadratic optimization subproblems in
(\ref{P2}) by employing the Lagrange dual method. Once given the
Lagrange multipliers, we can obtain a unique closed-form solution for
the Lagrange function minimization problem. Additionally, the number
of dual variables is usually much lower than that of the primal
variables $\bm{x}$ in the quantized mMIMO regime. Accordingly, solving
the dual problem is more efficient than directly solving the optimal
primal problem.

For the problem at hand, the first step is to introduce the
nonnegative Lagrange multipliers $\boldsymbol{\lambda} =
[\lambda_{1},\ldots,\lambda_{K}]^{T}, \boldsymbol{\mu} =
[\mu_{1},\ldots,\mu_{N}]^{T}, \boldsymbol{\varrho} =
[\varrho_{1},\ldots,\varrho_{K}]^{T}, \boldsymbol{\delta} =
[\delta_{1},\ldots,\delta_{S}]^{T}, \boldsymbol{\Phi} =
\left[\boldsymbol{\phi}_{1},\cdots,\boldsymbol{\phi}_{S}\right]$
associated with the constraints of (\ref{P2}). For clarity, we denote
$\Theta =
[\boldsymbol{\lambda}^{T},\boldsymbol{\mu}^{T},\boldsymbol{\varrho}^{T},\boldsymbol{\delta}^{T},\mathrm{vec}(\boldsymbol{\Phi})^{T}]^{T}$. Then
we write the Lagrange function of problem (\ref{P2}) in
(\ref{lagrange}) as displayed at the bottom of this
page. Then the dual function for (\ref{lagrange}) can
  be expressed as
$f_{d}^{l}(\Theta) = \underset{\mathbf{x}}{\min}\quad\mathcal{L}^{l}(\mathbf{x},\Theta)$
and the corresponding dual problem is given by
\setcounter{equation}{15}
\begin{equation}\label{dual_max}
\underset{\Theta}{\max}\quad f_{d}^{l}(\Theta).
\end{equation}
Note that $f_{d}^{l}(\Theta)$ can be further decomposed into $N$ independent subproblems as
$\underset{x_{t}\in {\chi}_{t}}{\min} a(\Theta)|x_{t}|^{2} + \Re\left[\sum_{t=1}^{n}b(\Theta)x_{t}\right]$,
which admits the closed-form solution as
\begin{equation}\label{closed_form}
\hat{x}_{t}(\Theta) = \mathbb{P}_{\chi_{t}}\left[-{b^{*}\left(\Theta\right)}/{2a(\Theta)}\right],
\end{equation}
where $\mathbb{P}_{\chi_{t}}[\cdot]$ is the projection over the convex set $\chi_{t}$. The optimal Lagrange multipliers $\Theta^{\star}$ of the dual problem in (\ref{dual_max}) can be solved using the standard subgradient-based method such as the ellipsoid method in \cite{boyd}. Then the optimal primal solution of (\ref{P2}) is given by $\mathbf{x}^{\star}(\Theta^{\star})$.
\begin{table*}[htbp]
\small
    \caption{Computational Complexity Orders for Different Schemes}
    \label{complexity}
    \begin{center}
    \begin{tabular}{cc}
    \hline
    \text{Schemes} & \text{Computational complexity order per iteration}\\
    \hline
    \text{SHC scheme} & $\mathcal{O}((N^{3}S^{3}+N^{2}S^{4}+NS^{4}+S^{2}K^{2})\mathrm{log}(1/\epsilon)+S^{2}K+S(NK+K^{2}))$  \\
    \text{MM scheme} & $\mathcal{O}((4S^{2}K^{2}+4SK^{3})\mathrm{log}(1/\epsilon)+S^{2}K+SK^{2})$\\
    \text{ZF scheme} & $\mathcal{O}((N^{3}S^{3}+N^{2}S^{3}K+NS^{3}+S^{2}K^{2})\mathrm{log}(1/\epsilon)+S(K^{2}+NK)+K^{3})$\\
    \text{MRC scheme} & $\mathcal{O}((N^{3}S^{3}+N^{2}S^{3}K+NS^{3}+S^{2}K^{2})\mathrm{log}(1/\epsilon)+S(K^{2}+MK+NK))$\\
    \hline
    \end{tabular}
    \end{center}
    \end{table*}
\vspace{-4mm}
\subsection{Convergence Analysis of the Proposed RSSCA Algorithm}
\vspace{-1mm}
By averaging all the outputs corresponding to the previous feasible updates or objective updates, we can obtain the limiting point $\mathbf{x}$ alternatively. Nevertheless, it is tricky to prove that the limiting point is a stationary point of problem \eqref{P}. To overcome these challenges, we first introduce the following lemma to guarantee the convergence of the recursive approximation $\hat{r}^l_k$ and the surrogate function $\hat{f}_{k}^{l}(\bm{x})$.
\begin{lemma}\label{le:recurCov}
 If the step-size sequence $\{\alpha^{l}\}$ and $\{\beta^{l}\}$ satisfy the following three conditions
 \begin{enumerate}
\item   $\alpha^{l} \to 0, \sum_{l}\alpha^{l} = \infty, \sum_{l}(\alpha^{l})^{2} < \infty$,

\item $\beta^{l} \to 0, \sum_{l}\beta^{l} = \infty, \sum_{l}(\beta^{l})^{2} < \infty$,

\item $\lim_{l \to \infty} \alpha^{l}/\beta^{l} = 0$,
 \end{enumerate}
we almost surely have
\begin{align}
\lim_{l\to \infty}|\hat{r}_{k}^{l}(\mathbf{x}^{l}) - \overline{r}_{k}(\mathbf{x}^{l})| &= 0, \label{lemma1}\\
\lim_{l\to \infty}|\hat{f}_{k}^{l}(\mathbf{x}^{l}) - f_{k}(\mathbf{x}^{l})| &= 0, \label{lemma2}\\
\lim_{l\to \infty}\|\nabla \hat{f}_{k}^{l}(\mathbf{x}^{l}) - \nabla f_{k}(\mathbf{x}^{l})\| &= 0.\label{lemma3}
\end{align}

Moreover, consider a subsequence $\{\mathbf{x}^{{l}_{j}}\}_{j=1}^{\infty}$ converging to a limiting point $\mathbf{x}^{\star}$, and define
\begin{equation*}
\tilde{f}_{k}(\mathbf{x})\triangleq \gamma_k-\overline{r}_k(\mathbf{x}^{\star})+\Re[\nabla^{H}f_k(\mathbf{x}^{\star})(\mathbf{x}-\mathbf{x}^{\star})]+\tau_k\|\mathbf{x}-\mathbf{x}^{\star}\|^2.
\end{equation*}
Then we almost surely have
$\lim_{j \to \infty} \hat{f}_{k}^{{l}_{j}}(\mathbf{x}) = \tilde{f}_{k}(\mathbf{x}), \forall \mathbf{x}\in\chi$.
\end{lemma}

Lemma~\ref{le:recurCov} can be proven following an approach similar to
that in~\cite{ssca}. Thus, the details are omitted due to the limited
space. According to~(\ref{lemma1})-(\ref{lemma3}) in Lemma
\ref{le:recurCov}, we may infer that the objective value and the
gradient of the surrogate function $\hat{f}_{k}^{l}$ as well as the
recursive approximation $\hat{r}^l_k$ are unbiased estimates of
$f_{k}(\mathbf{x}^{l})$ and $ \overline{r}_{k}^{l}(\mathbf{x}^{l})$,
respectively. In the sequel, we briefly discuss the motivation for
some conditions on $\{\alpha^{l},\beta^l\}$. Firstly, condition 1) and
2) state that both the step-size $\alpha^{l}$ and $\beta^{l}$ satisfy
the diminishing rule, namely, do not decay to zero sharply. Without
loss of generality, the diminishing step-size rule has also been
proposed in many other stochastic optimization
solutions~\cite{ssca}. Secondly, condition 3) clarifies that the
diminishing speed of $\alpha^{l}$ is faster than that of $\beta^{l}$.
With Lemma \ref{le:recurCov}, we can have the following theorem.

\begin{theorem}
Suppose the above assumptions are satisfied. For any sequence
$\{\mathbf{x}^{{l}_{j}}\}_{j=1}^{\infty}$ converging to a limiting
point $\mathbf{x}^{\star}$, problem (\ref{P}) almost surely converges
to stationary point $\mathbf{x}^{\star}$ if the Slater condition is
satisfied.
\end{theorem}



\vspace{-5.5mm}
\subsection{Computational Complexity}
\vspace{-2mm}
In this subsection, we compare the computational complexity of the proposed RSSCA algorithm to that of the following baseline schemes.
\vspace{-2mm}
\begin{itemize}
\item \textbf{Baseline 1-magnitude maximization (MM) scheme:} This is obtained by selecting a set of codewords according to the maximum SNR criterion \cite{MM}.

\item \textbf{Baseline 2-zero-forcing (ZF) scheme:} This is obtained by fixing the ZF digital combiner \cite{b1}.

\item \textbf{Baseline 3-maximum ratio combining (MRC) scheme:} This is obtained by fixing the MRC digital combiner \cite{two_stage}.
\end{itemize}
\vspace{-1mm} For simplicity, we assume that $M\gg
  N\geq S\geq K$. We first analyze the computational complexity of the
  proposed RSSCA algorithm. Then, the computational complexity order
  of the other schemes can be obtained similarly.
The complexity of the proposed RSSCA algorithm is
  dominated by the calculation of the gradient $\nabla
  f_{k}(\mathbf{x}^{l})$ and the quadratic optimization subproblems in
  (\ref{P2}) and (\ref{P3}), which is elaborated on below. Calculating
  $\nabla f_{k}(\mathbf{x}^{l})$ requires $\mathcal{O}(K(K\!+\!NS\!+\!S^{2}\!+\!KS))$
  floating point operations (FPOs). For the given Lagrange multipliers,
  calculating the closed-form primal solution in (\ref{closed_form})
  needs $\mathcal{O}(K+NS+S^{2}+KS)$ FPOs. Using the ellipsoid method,
  the number of iterations required for achieving a given convergence
  accuracy $\epsilon$ for the dual problem $f_{d}^{l}(\Theta) =
  \underset{\mathbf{x}}{\min}\mathcal{L}^{l}(\mathbf{x},\Theta)$ is
  $\mathcal{O}((2K+N+S+NS)^{2}\mathrm{log}(1/\epsilon))$
  \cite{boyd}. Hence, the per-iteration calculation needs
  $\mathcal{O}((K+NS+S^{2}+KS)(2K+N+S+NS)^{2}\mathrm{log}(1/\epsilon)+K(K+NS+S^{2}+KS))$
  FPOs.

In Table I, we summarize the computational complexity
  orders of the different schemes.  For the ZF scheme this is
  $\mathcal{O}((N^{3}S^{3}+N^{2}S^{3}K+NS^{3}+S^{2}K^{2})\mathrm{log}(1/\epsilon)+S(K^{2}+NK)+K^{3})$
  because in addition to the calculation of the gradient and the
  quadratic optimization subproblems it requires the calculation of
  inversion, when adopting the ZF digital combiner. Furthermore, the
  complexity order of the MRC scheme can be analyzed in a similar
  way. Although the MM scheme imposes a slightly lower computational
  complexity, its performance is in general much worse than that of
  our proposed SHC scheme. Consequently, our proposed SHC scheme
  strikes a better performance vs.  complexity trade-off.
\vspace{-4mm}
\subsection{Implementational Considerations}
\vspace{-1.5mm} Next, we investigate the pilot overhead and quantify
the advantages over the ICSI-based scheme as follows. Specifically,
the overall pilot overheads of the SHC and ICSI-based schemes are
$O(KML_c)$ and $O(KML_fL_s)$, respectively, where $L_c$ is the number
of frames required by the proposed algorithm. Hence, the SHC scheme
advocated reduces the pilot overhead by a factor of $\frac{L_c}{L_f
  L_s}$. Then, we compare the robustness of different schemes in terms
of the feasible probability of the average rate constraint~\eqref{P1A}
when $M=64$, $S=12$, and $K=12$. By averaging over 500 independent
channel realizations, the feasible probability of the proposed SHC and
the ICSI-based schemes is $90.05\%$ and $54.53\%$, respectively. The
reason is that the performance of the ICSI-based scheme is crucially
dependent on the CSI estimated at each time slot, and having ICSI
errors caused by the delay are inevitable in practice due to the
limited training resources. However, the acquisition of CSI in the SHC
scheme is performed at each frame with higher accuracy. The delayed
CSI is generated based on the autoregressive model
of~\cite{binary}. In short, the proposed SCH scheme is more robust to
the CSI signaling delay than the ICSI-based scheme due to the
sophisticated stochastic design and owing to the reduction of
signalling bits.

\vspace{-4.8mm}
\section{Simulation Results}
\vspace{-1.8mm} In this section, we numerically evaluate the performance
of our proposed scheme and glean useful insights. We consider the
scenario of $M = 64$ receive antennas, $S = 12$ RF chains, and $N =
16$ DFT codewords. The maximum transmit power at each user is
$P_{k}^{\mathrm{max}} = 10$ dBm. We set the same
  average target rate for all users, namely $\gamma_{k} = 1$ bps/Hz.
The BS's coverage has a radius of $200$ m and $K=12$ users are
randomly located. Additionally, the
  channel between user $k$ and the BS is generated according to the
  extended Saleh-Valenzuela geometric model using a half-wavelength ULA
  \cite{ssca}.
We use the large-scale pathloss model $30.6 + 36.7
\mathrm{log}_{10}(d_{k})$ \cite{pathloss} to characterize the average
channel gain, where $d_{k}$ is the individual distance between user
$k$ and the BS in meters.  The initial parameters are set as $\beta =
1/(1+l)^{2/3}$ and $\alpha = 5/(5+l)$,
respectively. Four schemes are included as
  benchmarkers: 1) MM scheme; 
  $\mathbf{C}$ concentrating on choosing the beams with maximum
  magnitude \cite{MM}; 2) random scheme, which is obtained by randomly
  associating each RF chain with the specific beam; 3) ZF scheme; 4)
  MRC scheme.

\begin{figure}[t]
		\centering
		\includegraphics[width=6cm]{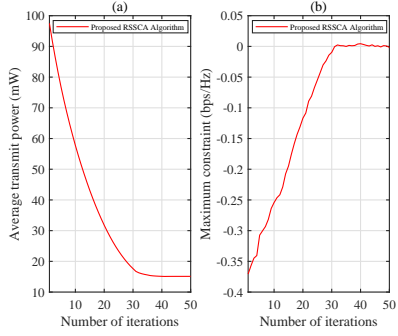}
		\caption{Convergence behavior of the proposed RSSCA algorithm with $M=64$ antennas, $K=12$ users, $S=12$ RF chains, $N = 16$ DFT codewords and $q=3$ quantization bits: (a) average transmit power; (b) maximum constraint.}
        \label{fig:converge}
\end{figure}

In Fig.~\ref{fig:converge}, we show the objective function (average
transmit power) and the maximum constraint function (target average
rate) versus the number of iterations for $q = 3$ quantization bits, respectively. We can see that
the proposed RSSCA algorithm converges within a few iterations, while
all the target average rates are strictly satisfied with high
accuracy.

\begin{figure}[h]
		\centering
		\subfigure{
			\begin{minipage}[t]{0.5\textwidth}
				\centering
				\includegraphics[width=6cm]{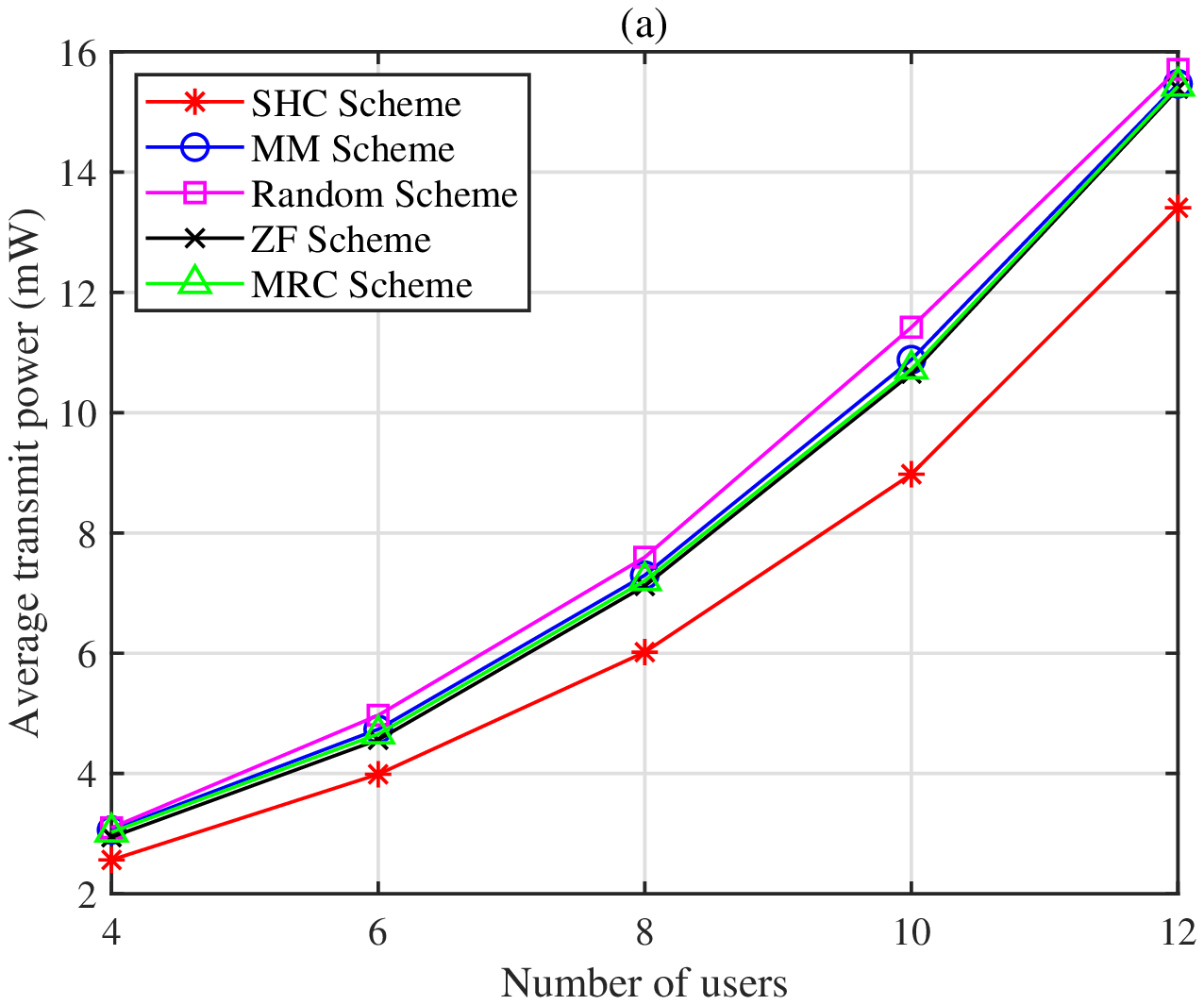}\label{fig:compareK}\hspace{-2cm}
                \includegraphics[width=6cm]{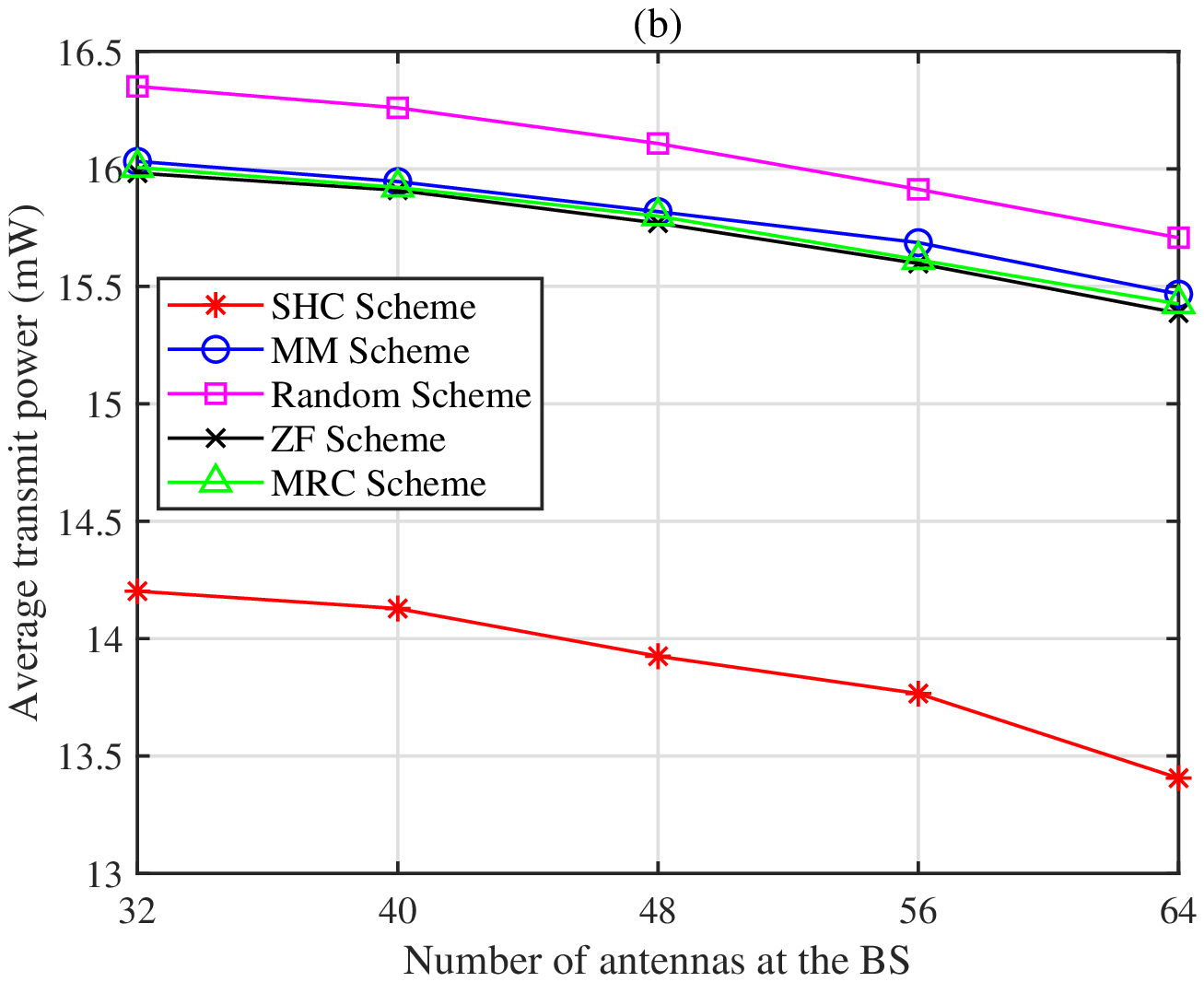}\label{fig:compareM}\hspace{-2cm}
                \includegraphics[width=6cm]{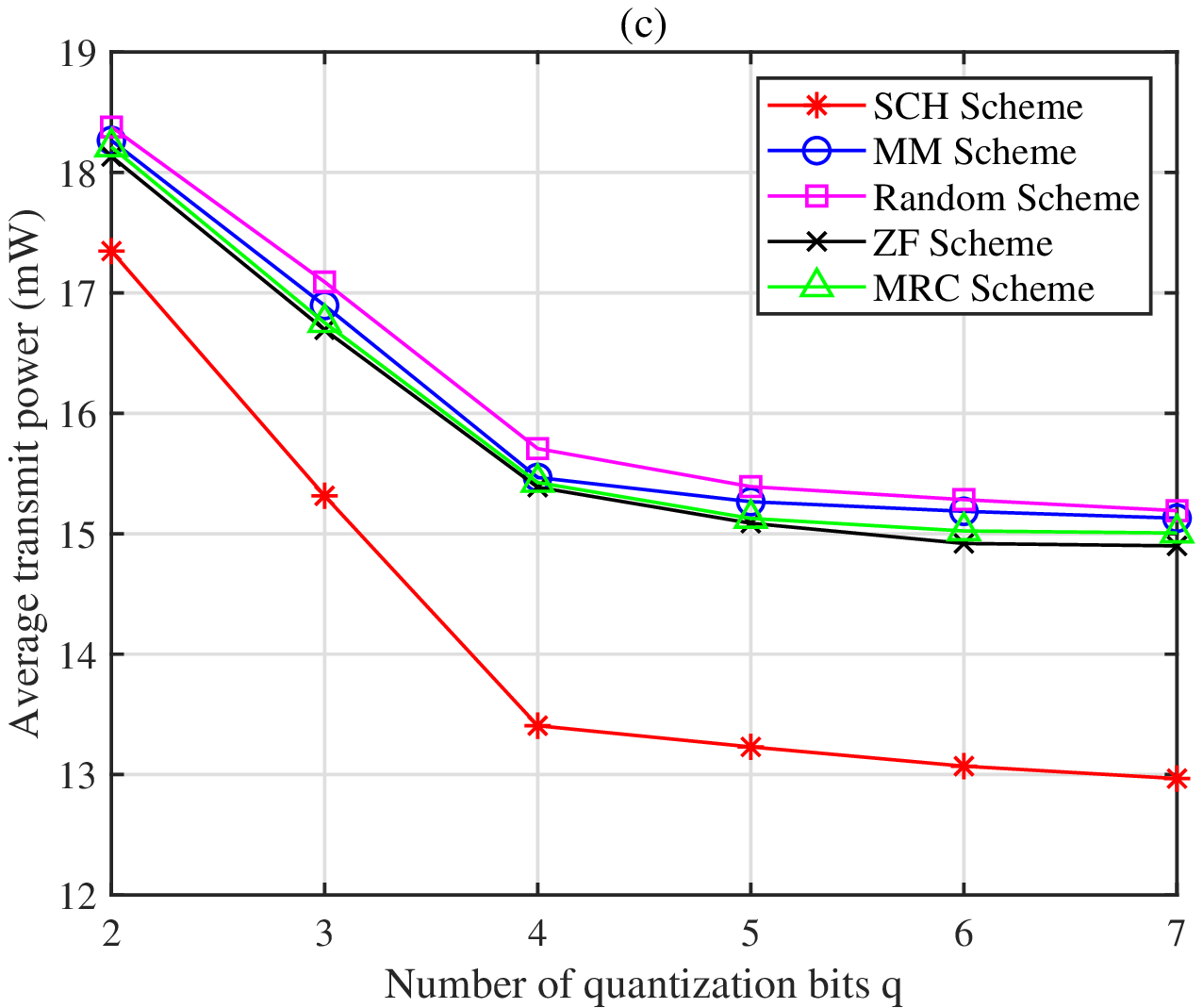}\label{fig:comparebit}
			\end{minipage}
		}
        \caption{(a) Average transmit power versus the number of
          users with $M=64$ antennas and 4-bits ADCs; (b) average transmit power versus the number of
          antennas at the BS with $K=12$ users and 4-bits ADCs; (c) average transmit
            power versus the number of quantization bits $q$ with $M=64$ antennas and $K=12$ users.}
        \label{fig:compare}
\end{figure}
In Fig. \ref{fig:compare} (a), we plot the average transmit power
versus the numbers of users $K$ for $q = 4$ quantization bits. It can be seen that the performance
of the proposed SHC scheme is superior to that of all the other
competing schemes. When the number of users increases, the performance
gap between the proposed SHC scheme and these other schemes is
widened, which also demonstrates the necessity of joint power control
and hybrid combiner design for the quantized mMIMO system. Thus, it
appears that for the quantized mMIMO relying on limited resources,
our proposed scheme is particular appealing from an optimum resource
allocation perspective. Fig. \ref{fig:compare} (b) depicts the
average transmit power versus the number of BS antennas $M$ with 4-bits ADCs. We
observe that the proposed SHC scheme significantly outperforms all
the other competing schemes, as expected. It is interesting to note
that the benefit of increasing the number of BS antennas is
more pronounced for our proposed scheme. The reason is that our
proposed scheme can exploit the difference in channel quality among
links for mitigating the multiuser interference, thereby
supporting more favorable uplink transmission in a cost-effective
manner.
Fig. \ref{fig:compare} (c) plots the system's average
transmit power versus the number of quantization bits $q$. It shows
that the average transmit power is monotonically decreasing as the
number of quantization bits increases. Furthermore, the proposed SHC
scheme outperforms all the other competing schemes in all
quantization scenarios, especially for more than 3 quantization
bits.
\vspace{-5.5mm}
\section{Conclusion}
\vspace{-2mm}
A novel SHC scheme was conceived for the uplink of
quantized mMIMO systems for minimizing the transmit power consumption
under an average rate constraint. We proposed a RSSCA algorithm to
find stationary solutions of the nonconvex stochastic optimization
problem. We demonstrated that the proposed SHC scheme
outperforms the benchmarkers.

\end{document}